# Anomalously Abrupt Switching of Ferroelectric Wurtzites

Keisuke Yazawa[1,2]\*, John Hayden[3], Jon-Paul Maria[3], Wanlin Zhu[3], Susan Trolier-McKinstry[3], Andriy Zakutayev[1], Geoff L. Brennecka[2]\*

[1]Materials Science Center, National Renewable Energy Laboratory, Golden, Colorado 80401, United States.

[2]Department of Metallurgical and Materials Engineering, Colorado School of Mines, Golden, Colorado 80401, United States.

[3]Department of Materials Science and Engineering and the Materials Research Institute, The Pennsylvania State University, University Park, Pennsylvania 16802, United States.

\*Corresponding author. Email: Keisuke.Yazawa@nrel.gov, geoff.brennecka@mines.edu

## Abstract

Ferroelectric polarization switching is one common example of a process that occurs via nucleation and growth, and understanding switching kinetics is crucial for applications such as ferroelectric memory. Here we describe and interpret anomalous switching dynamics in the wurtzite nitride thin film ferroelectrics $Al_{0.7}Sc_{0.3}N$ and $Al_{0.94}B_{0.06}N$ using a general model that can be directly applied to other abrupt transitions that proceed via nucleation and growth. When substantial growth and impingement occur while nucleation rate is increasing, such as in these wurtzite ferroelectrics under high electric fields, abrupt polarization reversal leads to very large Avrami coefficients (e.g., $n = 11$), inspiring an extension of the KAI (Kolmogorov-Avrami-Ishibashi) model. We apply this extended model to two related but distinct scenarios that crossover between (typical) behavior described by sequential nucleation and growth and a more abrupt transition arising from significant growth prior to peak nucleation rate. This work therefore provides more complete description of general nucleation and growth kinetics applicable to any system while specifically addressing both the anomalously abrupt polarization reversal behavior in new wurtzite ferroelectrics.

## Introduction

The kinetic description of phase transformations via nucleation and growth represents one of the most fundamental concepts across many branches of science. More than 80 years ago, the equations now collectively referred to as Johnson-Mehl-Avrami-Kolmogorov (JMAK) were developed to describe and interpret the kinetics of isothermal crystallization via nucleation and growth[1–5]. The predictable sigmoidal transition from state A to state B enables descriptions of inherently local phenomena of nucleation and impingement-limited growth from global measurements; this has been applied to a broad class of problems including microstructure development in high strength steels[6], cloud formation and weather forecasting[7,8], pharmaceutical manufacturing[9], pollution adsorption[10], and many others. In applying the same approach to ferroelectric switching, Ishibashi expressed what is now referred to as the Kolmogorov-Avrami-Ishibashi (KAI) model[11]:

$$f = 1 - \exp\left\{-\left(\frac{t}{t_0}\right)^n\right\} \qquad (1)$$

where $f$ is the volume fraction switched at time $t$, $t_0$ is the characteristic time for polarization evolution, and $n$ is the Avrami exponent.

While Ishibashi explicitly addressed only two simple cases for nucleation during ferroelectric switching—either pre-existing nuclei (Avrami exponent, $n$, equals growth dimensionality, $D$) or a constant continuous nucleation rate as a result of concurrent nucleation and growth ($n = 1 + D$)[12]—the KAI model has been used extensively in the past five decades to interpret ferroelectric switching data in terms of nucleation rates and growth velocities, in many cases without explicit acknowledgement of the boundary conditions (the nucleation scenarios) associated with the original derivation[13–16]. Indeed, in treating nucleation and growth as independent processes, there are three possible sequences (Fig. 1): (A) the nucleation-limited switching (NLS) model with no significant domain growth, (B) the KAI model governed by domain growth from pre-existing nuclei followed by impingement, and (C) a scenario in which nucleation rate peaks significantly after the onset of growth and impingement. Tagantsev, et al. rationalized $n < 1$ via the NLS model[17–20] represented schematically in Fig. 1a. Such behavior, seen exclusively in ferroelectric switching, has been ascribed to a broad distribution of coercive fields resulting from a distribution of polarization directions relative to the applied electric field and/or a broad distribution of pinning depths for interface motion[21]. Scenarios consistent with the KAI model, where nucleation dominates early-time processes followed by domain growth and impingement in 1, 2, or 3 dimensions, are represented in Fig. 1b. Several scanning probe studies[15,22–25] have reported time-varying nucleation rates inconsistent with Ishibashi's assumptions, but no work has yet addressed the scenario shown in Fig. 1c in which significant growth occurs prior to and throughout a peak in nucleation rate.

As Cahn noted in his classic rederivation[26], the JMAK model itself has no restrictions against time-varying nucleation and growth rates, but $n$ is often overinterpreted as *necessarily* being equal to either the dimensionality of growth ($n = D$) or one plus the dimensionality of growth ($n = 1 + D$). A decaying nucleation rate following the Jacobs-Thompkins equation[27] has been shown to be consistent with a mixed scenario ($D \leq n \leq 1 + D$) and is commonly applied to solid state chemistry and crystal growth[28,29], but examples requiring a more nuanced description of nucleation contributions have been limited,[25,30–33] such that the $D \leq n \leq 1 + D$ interpretation is pervasive. Elaboration of nucleation and growth kinetics beyond $D \leq n \leq 1 + D$ is required to capture the scenario shown in Fig. 1c.

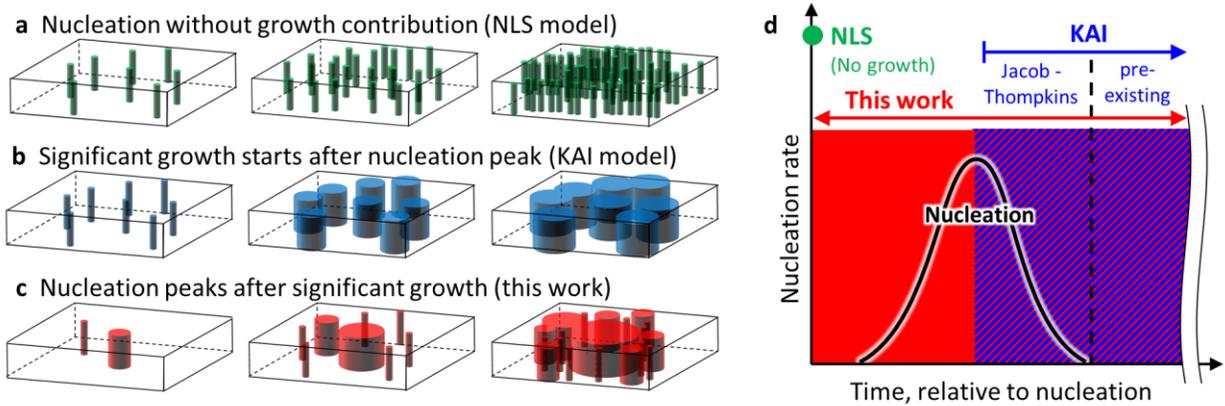

**Fig. 1 Prior work has successfully described the ferroelectric switching process according to** (**a**) nucleation-limited switching with no significant domain growth, (**b**) the KAI model in which nucleation can be described by a population of pre-existing nuclei and/or a (Jacob-Thompkins) decreasing nucleation rate followed by growth and impingement. (**c**) Here, the KAI model is extended to include scenarios in which nucleation rate peaks significantly after the onset of growth. This is done by (**d**) describing all nucleation processes according to a power law distribution and relaxing the assumptions that restrict the KAI model to specific cases in which nucleation rate peaks prior to the onset of significant growth.

We show here that the switching of $Al_{0.7}Sc_{0.3}N$ and $Al_{0.94}B_{0.06}N$ wurtzite ferroelectrics can contradict the $D \leq n \leq 1 + D$ interpretation of the Avrami exponent, as this would require growth in as many as 10 dimensions to fit measured data, which is clearly non-physical. Moreover, by extending the KAI model to account for a time-varying nucleation rate described with a derivative of a sigmoid function (Fig. 1d), both this abrupt response and those well-described by the KAI model can be fit and physically interpreted, regardless of whether the transformation in question is ferroelectric switching, solidification, cloud formation, pharmaceutical manufacturing, or any other example of nucleation and growth.

## Results

### Anomalous switching kinetics and model extension

Wurtzite nitride ferroelectrics often exhibit square hysteresis loops[34–37], a characteristic known to relate to switching kinetics[38]. Fig. 2a plots the normalized polarization ($P/P_r$, where $P_r$ is remanent polarization) *vs.* normalized field ($E/E_c$, where $E_c$ is coercive field) for the prototype ferroelectrics (blue) single crystal $LiTaO_3$ and (green) polycrystalline thin film $PbZr_{0.53}Ti_{0.47}O_3$ (PZT) against polycrystalline (orange) $Al_{0.7}Sc_{0.3}N$ and (red) $Al_{0.94}B_{0.06}N$ thin films (see Methods)[37,39]. As shown in Fig. 2b, the polarization evolution with time for each of these samples results in excellent fits to the data with Avrami exponents $n = 2$ for $LiTaO_3$ (traditional KAI)[40], $n < 1$ for PZT (NLS)[19], and $n > 4$ for $Al_{0.7}Sc_{0.3}N$ and $Al_{0.94}B_{0.06}N$. Forcing $n = 2$ or 4 (Fig. 2c and Fig. 2d) for $Al_{0.7}Sc_{0.3}N$ and $Al_{0.94}B_{0.06}N$ clearly results in unacceptable fits, but $n > 4$ defies direct physical interpretation using the common $D \leq n \leq 1 + D$ modality of the KAI model.

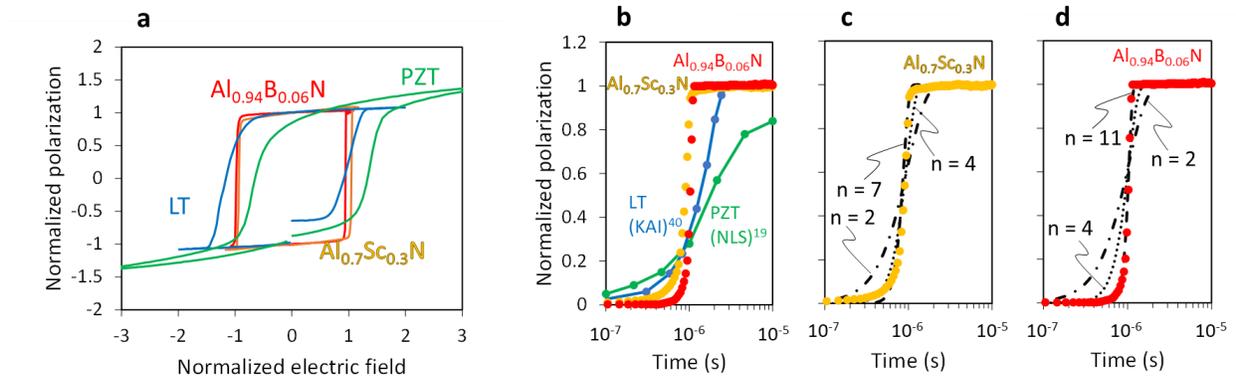

**Fig. 2 New wurtzite ferroelectrics require an update to switching models.** (**a**) Recently discovered wurtzite nitride ferroelectrics exhibit hysteresis loops that are far more square than those of prototype ferroelectrics such as single crystal LiTaO$_3$ and thin film PZT. (**b**) While polarization evolution of a LiTaO$_3$ crystal and PZT thin film can be fit with KAI and NLS interpretations, respectively[19,40], the Avrami exponent required to fit the data for Al$_{0.7}$Sc$_{0.3}$N and Al$_{0.94}$B$_{0.06}$N has no physical interpretation according to existing models. (**c**) Avrami exponent to fit Al$_{0.7}$Sc$_{0.3}$N data is 7. (**d**) Avrami exponent to fit Al$_{0.94}$B$_{0.06}$N data is 11.

We therefore revisit the classic JMAK model without forcing adherence to either the coalescence-dominated KAI model ($1 < n < 4$) or the nucleation-limited NLS model ($n < 1$). Based on the assumption of an infinite system with growth ending by impingement, the transformed fraction as a function of time $t$ can be represented as[3],

$$f = 1 - \exp\left[-\int_0^t \dot{N}(\tau) V(t,\tau) d\tau\right] \quad (2)$$

where $\dot{N}$ is the nucleation rate, $V$ is the volume of each transformed region, and $\tau$ is the nucleation time.

To incorporate a nucleation rate peak as discussed above (Fig. 1d), we employ a saturation function including the $m^{th}$ order nucleation rate as a function of time:

$$\dot{N}(\tau) = \alpha N(\infty) m \tau^{m-1} \exp(-\alpha \tau^m) \quad (3)$$

where $N(\infty)$ is the saturated density of nuclei and $\alpha$ is a constant determining the time of the peak position. This time-varying nucleation rate captures both the incubation time and non-linear increase of nucleation[41,42], in contrast to the classic KAI theory. Among the many possible non-linear functions to express the nucleation behavior, the $m^{th}$ order nucleation rate has been suggested for broadening the interpretation of kinetics in non-isothermal phase transition kinetics.[43] By plugging this form of the nucleation rate into equation (2) and assuming isotropic 2-dimensional growth (see Methods), the volume fraction transformed (or switched, when applied to polarization reversal) in a thin film can be expressed as

$$f = 1 - \exp\left[-2\pi d v^2 N(\infty) \alpha m \int_0^t \tau^{m-1}(t-\tau)^2 \exp(-\alpha \tau^m) \, d\tau\right] \quad (4)$$

The time evolution of switching can then be fitted using parameters $v$, $\alpha$ and $m$, which describe the growth speed, nucleation rate peak position and shape, respectively. As shown in equation (4), it is not necessary to decouple the growth velocity from the saturated density of nuclei, so a coupled variable $vN(\infty)^{1/2}$ is employed for calculation. A qualitative exploration of the effects of each of these parameters independently offers helpful insights. Changes in growth speed shift the polarization evolution curve along the logarithmic time axis; this is similar to the effects of an electric field on $t_0$ in the classic KAI model. The constant $\alpha$ determines the nucleation peak position. The exponent $m$ stands for the $m^{th}$ order nucleation rate, which represents the experimentally observed nucleation rate peak, not considered in previous KAI and KAI-derived models.[13,44,45]

**Electric field dependent nucleation vs growth competition**

Experimental results of electric field dependence of polarization evolution for $Al_{0.7}Sc_{0.3}N$ and $Al_{0.94}B_{0.06}N$ thin films validate the extended model (Fig. 3). The polarization evolution curves shift to faster times with larger applied electric field (Fig. 3a and b), conceptually consistent with Merz's law[46] and in a manner similar to comparable studies on perovskite oxide ferroelectric thin films[19,25,47]. In addition to the shift, the slope of the polarization increase steepens with increasing applied electric field for both materials. This slope change is also seen in conventional ferroelectrics[19,47], though only the recent study from Nath, et al., has reported n values $> 3$[25]. In contrast, the KAI fitted Avrami exponents of $Al_{0.7}Sc_{0.3}N$ and $Al_{0.94}B_{0.06}N$ films reach 7 and 11, respectively, at high electric fields (Fig. 2c, 2d).

Our extended model describes both abrupt transitions (n > 4) seen in the wurtzite ferroelectrics and conventional KAI transitions (1 < n < 4). The black solid lines in Fig. 3a and b are numerically simulated curves (see Methods) based on the extended model for each applied electric field. The value of $m$ is set equal to 5 for $Al_{0.7}Sc_{0.3}N$ and 9 for $Al_{0.94}B_{0.06}N$ in the simulation. Note that the $m$ values represent the order of nucleation rate, different from the $n$ values that are fitted with the classic KAI model. Normalized nucleation rate curves used for the simulations, which are determined by $\alpha$ (see Methods), are shown in Fig. 3c and 3d. The coupled variable $vN(\infty)^{1/2}$ related to domain wall (growth) velocity is chosen to fit the experimental curves (see Methods). As noted earlier, the model assumes 2-dimensional growth, so without resorting to non-physical growth dimensionality, the extended model accounts for *and provides supplemental information about* the steep evolution of the polarization reversal process via the nucleation peak.

At higher electric fields, the nucleation peak (dotted line) occurs *after* the switching event is already complete; in other words, 2-dimensional lateral domain growth and coalescence occur while the nucleation rate is increasing. Simultaneous nucleation and growth produce the abrupt transition of the model, fitting the steep slope observed under high electric fields. Conversely, under lower electric fields, the nucleation rate peaks before significant polarization reversal has occurred, equivalent to the presence of preexisting nuclei before significant domain growth, precisely as assumed in the KAI model. In such cases, *any* value of $m$ produces the same overall result because the shape of the time-varying nucleation peak is irrelevant if growth is insignificant while nucleation is occurring. Thus, when nucleation peaks prior to the onset of significant growth, the extended model collapses to precisely the delta ($n = D$) or Jacobs-Thompkins decay ($n = 1 + D$) of the KAI model. Between these extremes of complete transition prior to peak nucleation and peak nucleation well before substantial growth, there is a gradual change in the slope of the polarization reversal as there is a crossover in the relative positions of the nucleation peak and the steepest rise of the polarization curve.

The crossover can be visualized in terms of characteristic times for both nucleation and growth. From the fitted $\alpha$ and $vN(\infty)^{1/2}$, the nucleation peak time $\tau_{peak}$ and average impingement time $\bar{t}_{imp}$ due to domain growth are calculated (see Methods). Each experimental result can be fit with a range of $\alpha$ (or $\tau_{peak}$) and $vN(\infty)^{1/2}$ (or $\bar{t}_{imp}$); i.e., different pairs of $\tau_{peak}$ and $\bar{t}_{imp}$ can produce the same simulation curve (Fig. S1, ESI). Fig. S2 (ESI) shows the fittable range of $\tau_{peak}$ and $\bar{t}_{imp}$ for each experimental result. A slower $\tau_{peak}$ needs a faster $\bar{t}_{imp}$ to fit the experimental curves, and such fits agree well at higher electric fields, while faster $\tau_{peak}$ and slower $\bar{t}_{imp}$ pairs fit better for lower electric fields. The condition $\tau_{peak} < \bar{t}_{imp}$ means that significant growth and impingement happen after peak nucleation, corresponding to the assumptions of the traditional KAI model, but $\tau_{peak} > \bar{t}_{imp}$ is not covered in earlier models. Our extended model covers both regions. Domain wall velocities calculated from the fitted range of $\bar{t}_{imp}$ (or $vN(\infty)^{1/2}$) can have reasonable values (less than the speed of sound in AlN[48]) at any possible nucleation density from a single nucleation site in a 1960 μm$^2$ device measured here (50 μm diameter) to a nucleation site for each unit cell (Fig. S3 and S4, ESI).

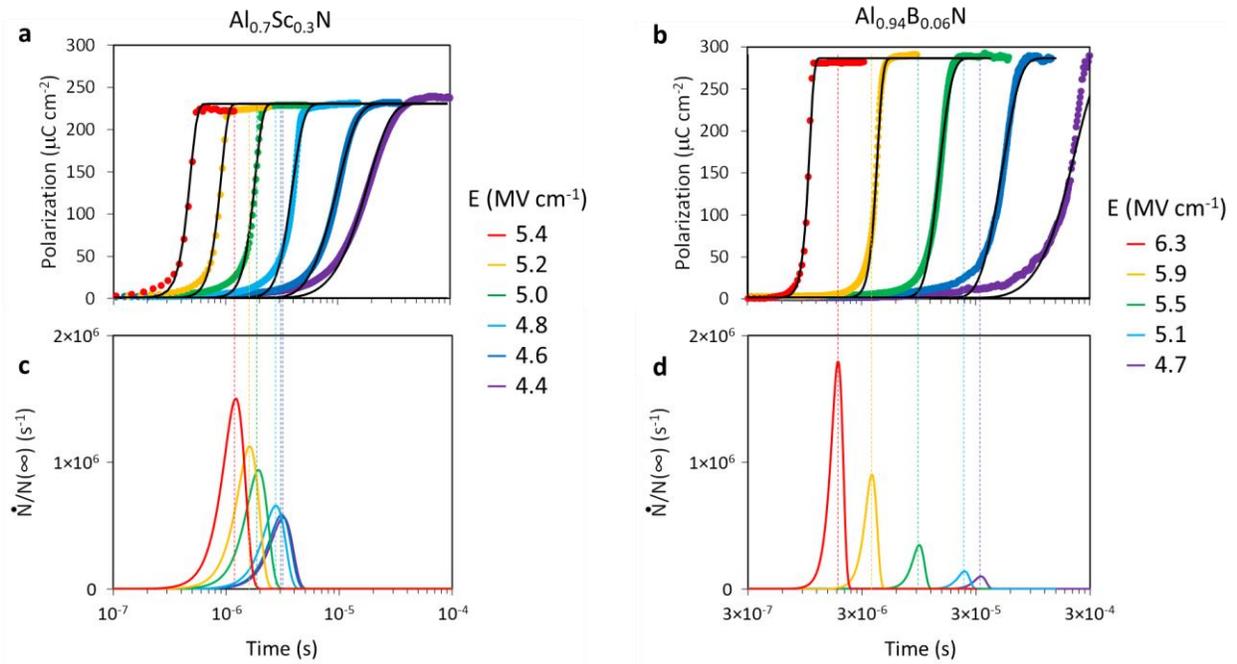

**Fig. 3 Evolution of switched polarization under various driving electric fields.** (**a**) Electric field dependence of the polarization vs. time curves for $Al_{0.7}Sc_{0.3}N$ and (**b**) for $Al_{0.94}B_{0.06}N$. Colored data points in the polarization – time curves represent measured data; black lines represent simulated results using the extended KAI model with $m = 5$ for $Al_{0.7}Sc_{0.3}N$ and $m = 9$ for $Al_{0.94}B_{0.06}N$. Normalized nucleation rate curves used for the simulation for (**c**) $Al_{0.7}Sc_{0.3}N$ and (**d**) $Al_{0.94}B_{0.06}N$.

**Cycle dependent nucleation vs growth competition**

The need to extend the classic KAI model to address abrupt switching and competition between nucleation vs. growth dominance is further validated by a closer look at the previously-described wakeup process of $Al_{0.94}B_{0.06}N^{49}$. The $Al_{0.94}B_{0.06}N$ film examined here exhibits comparable wakeup behavior (Fig. S5, ESI). The successive PUND sequences are applied through the top electrode (Fig. S6, ESI). Polarization evolution curves show no switching in the first cycle, partial switching at lower fields for the 2nd cycle, and gradually later initiation of a more abrupt transition for cycles > 3 (Fig. 4a). As the slope of the curve gets steeper, the KAI *n* also increases from 2 to 11 (Fig. 2d). The fittable range of $\tau_{peak}$, represented as brown bars in Fig. 4b, is extracted from the polarization evolution and shifts slower as cycle number increases. The fitting quality map as a function of fitting parameters $\tau_{peak}$ and $\bar{t}_{imp}$ is found in Fig. S7 (ESI), showing the transition from $\tau_{peak} < \bar{t}_{imp}$ (KAI) to $\tau_{peak} > \bar{t}_{imp}$ (beyond KAI).

The change in $\tau_{peak}$ with cycling is key to the wakeup behavior and is verified through the evolution of N-polar and metal-polar mixing. These films were N-polar as deposited; the PUND pre-pulse to the top electrode is intended to set the sample to metal-polar, then application of a P pulse switches the film (back) to the N-polar (downward polarization) state. Piezoelectric $d_{33}$ measurements prior to each P pulse confirm the polarity and show an intermediate-magnitude response consistent with incomplete poling in the film (see Methods and Fig. S6, ESI). The magnitude of $d_{33}$ measured prior to the P pulse gradually increases with cycle number as overlaid in Fig. 4b, direct evidence of the existence and decreasing volume fraction of N-polar regions even before the P pulse application. Such N-polar regions function as pre-existing nucleation sites for switching to the N-polar state during a P pulse, hence the earlier $\tau_{peak}$. With increasing cycles, the volume fraction of residual N-polar regions decreases and $d_{33}$ approaches that of the as-deposited film (schematic in Fig. 4e). Observation of the chemically etched film (see Methods) suggests that the pre-existing N-polar regions are likely located at the film/bottom electrode interface (Fig. S8, ESI).

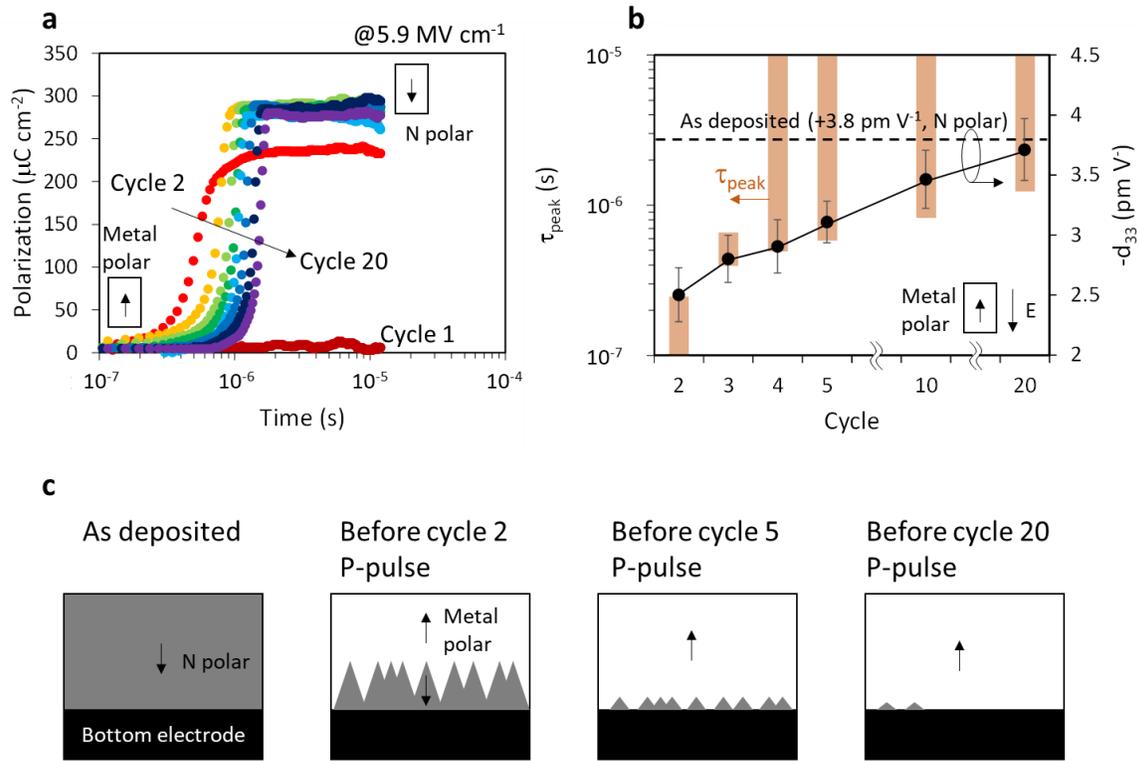

**Fig. 4 Kinetics during wakeup process in Al$_{0.94}$B$_{0.06}$N.** (**a**) Polarization evolution obtained from P - U pulses. Curve shift and slope change are observed. (**b**) Fittable range of $\tau_{peak}$ from polarization evolution curves using extended KAI model (brown bars) and piezoelectric coefficient d$_{33}$ (black dots) for each cycle before switching dynamics measurement. Error bars in d$_{33}$ represent a standard deviation of successive 20 measurements. At fewer cycle, d$_{33}$ decreases (mixed polarity) and $\tau_{peak}$ gets earlier (instant nucleation). (**c**) Schematic of domain evolution with number of cycles. Opposite polarity mixing seen at 2nd cycles relieves with cycles. The pre-existing opposite polarity works as nuclei, corresponding to early $\tau_{peak}$ at fewer cycles.

## Discussion

Transitions that occur via nucleation and growth can be dominated by either growth speed or nucleation rate. When the activation field for nucleation is smaller than that for growth, abrupt transitions can defy the traditional KAI model. It is noteworthy that the model presented in this study illuminates the characteristic times for nucleation ($\tau_{peak}$) and growth ($\bar{t}_{imp}$), which are often convoluted and difficult to separate. The discussion here uses the language of ferroelectric switching, but just like the JMAK foundation on which it is based, the model and interpretation are equally valid for any form of phase transition that meets the underlying assumptions and boundary conditions; for the examples described here, those are simply isotropic 2-dimensional growth within an infinite system with growth ending by impingement.

It has been reported that the activation field of nucleation is smaller than that of growth for switching in $Pb(Zr,Ti)O_3$[23], and peaking nucleation rates have also been reported in both $Pb(Zr,Ti)O_3$[15,22,23] and $HfO_2$[24]. Indeed, it has commonly been seen that $n$ increases with higher electric field in perovskite ferroelectric switching[19,25,47], which can be interpreted as a sign of approaching the crossover point. Avrami exponents $n > 4$ were reported in one such study but mechanistic discussion was limited[25]. To observe $n > 4$, the crossover point must be in the measurement time range. Although the origin of the slow crossover point in the wurtzite ferroelectrics is not known, nucleation may be hindered by the large interfacial energy of head-to-head polar boundaries due to the large spontaneous polarization (~110 µC cm$^{-2}$ for $Al_{0.7}Sc_{0.3}N$ and ~140 µC cm$^{-2}$ for $Al_{0.94}B_{0.06}N$), and/or a strong bond to be broken to initiate nucleation. The fact that the crossover point of $Al_{0.7}Sc_{0.3}N$ (2 µs) is earlier in time than $Al_{0.94}B_{0.06}N$ (5 µs) is consistent with the head-to-head interfacial energy discussion attributed to the large remanent polarization. This could also explain why, despite significantly faster switching of perovskite oxide films[50,51], only a single study has observed similar behavior[25]. Slow nucleation coupled with rapid transition is analogous to supercooling a liquid to induce fast crystallization[52] or dendritic growth, though without diffusion-limited kinetics. Thus, the phenomenon can be generalized to any nucleation-growth scenario with slow nucleation caused by a large nucleation energy barrier or limited diffusion.

Few studies of nucleation-growth kinetics have reported $n > 4$. Toth reported $n$ values as high as 4.87 for the crystallization of a quenched and reheated Metglas 2605 alloy (iron-based alloy with silicon and boron additives) and interpreted this value as indicative of simultaneous nucleation and growth with a nucleation rate that increased with time[30]. Pradell et al. reported values of $n$ that slightly exceeded 4 only in the very early stages of the primary crystallization of $Fe_{78}Si_{22}$ from amorphous $Fe_{73.5}Si_{17.5}CuNb_3B_5$[31]. Jeon et al. demonstrated two step phase transition kinetics with $n = 5.75$ for the first step of the $Ge_2Sb_2Te_5$ crystallization process. Those papers briefly noted the importance of nucleation contribution to the $n > 4$ values, and an analysis of a nucleation effect on the $Ge_2Sb_2Te_5$ crystallization kinetics exists[53], but no rational explanation and formulation of this abrupt transition had been explored. Recent work from Nath et al. reported n > 4 values for select regions of epitaxial PZT thin films via scanning probe and connected the large $n$ values with a lack of existing nucleation sites, but further mechanistic discussion was limited[25]. The extended model demonstrated in this work rationalizes the phenomena of such rapid transitions via characteristic nucleation and growth times and is also applicable to the nucleation–growth mechanism in crystallization processes beyond ferroelectric switching.

## Conclusions

In summary, we measured rapid polarization reversal in new wurtzite ferroelectrics, which manifests as large Avrami exponents that are non-physical based upon the common (over)interpretation of the traditional KAI model. An extension to the KAI model is proposed that incorporates non-linear time-varying nucleation rates such as those that have been experimentally reported. This extended model captures the switching behavior of both the new wurtzite ferroelectrics as well as conventional ferroelectrics, highlighting the distinct nucleation and growth events and competition between them. When peak nucleation occurs prior to significant growth, our extended KAI model collapses to the original KAI model. However, the extended model also covers regimes where significant growth occurs while the nucleation rate is still increasing, corresponding to a larger activation field for growth than for nucleation. This work explains both the abrupt switching behavior and wakeup behavior of the new wurtzite nitrides using an extension of the classic KAI model that offers a framework for better understanding nucleation and growth kinetics in general.

## Methods
### Kinetics model

Ishibashi approximated nucleation using either (Fig. 1d, pre-existing) a delta function at $t = 0$ or (Fig. 1d, Ishibashi) a constant. Deutscher later extended this[29] to describe a decreasing nucleation rate with time (Fig. 1d, Jacob-Tompkins) using the Jacob-Tompkins equation[27]

$$N(\tau) = N(\infty)\{1 - \exp(-\alpha\tau)\} \quad (5)$$

namely,

$$\dot{N}(\tau) = \alpha N(\infty)\exp(-\alpha\tau) \quad (6)$$

where $N(\infty)$ is the saturated density of nuclei, and $\alpha$ is the constant nucleation rate at the beginning that determines the starting time for the nucleation rate decay. In these cases, the Avrami exponent corresponds to either the dimensionality of growth ($0 < n < 3$) when nucleation is no longer occurring or the growth dimension plus the time dimension, such that $1 < n < 4$.

To realize the peaked nucleation rate seen in the experimental results, a non-linear nucleation rate at the beginning with subsequent decay is employed, namely

$$N(\tau) = N(\infty)[1 - \exp(-\alpha\tau^m)] \quad (7)$$

The time derivative of the nucleation curve corresponds to the nucleation rate, which is expressed in equation (3) in the main text.

Incorporating this form of the nucleation rate into the original JMAK model, equation (2), the volume fraction transformed can be expressed as:

$$f = 1 - \exp\left[-\alpha N(\infty)m \int_0^t \tau^{m-1}\exp(-\alpha\tau^m)\,V(t,\tau)d\tau\right] \quad (8)$$

In isotropic 2-dimensional growth, which is a common assumption in high aspect ratio thin films(13), the volume of each nuclei is written as:

$$V(t,\tau) = 2\pi d v^2 (t - \tau)^2 \quad (9)$$

where $d$ is the film thickness. Thus, the volume fraction transformed in a thin film form can be expressed as equation (4) in the main text.

### Nucleation peak time and average impingement time

The nucleation rate peak time $\tau_{peak}$ can be derived simply from the derivative of the nucleation rate (equation (3)):

$$\frac{d^2N(\tau)}{d\tau^2}\bigg|_{\tau=\tau_{peak}} = 0 \tag{10}$$

A solution of the equation can be expressed as:

$$\tau_{peak} = \sqrt[m]{\frac{m-1}{m\alpha}} \tag{11}$$

The average impingement time can be elucidated from the fitted velocity $v$ and $N(\infty)$. In the case of 2-dimensional growth, it is convenient to consider the areal nucleus density $N(\infty)^{2D}$, which is expressed as:

$$N_{(\infty)}^{2D} = dN(\infty) \tag{12}$$

where $d$ is the thickness of the film. This represents the area density of pillar-like through-thickness nuclei, presuming significantly faster forward domain wall motion compared to lateral growth[46,54]. Under the assumption of uniformly distributed nuclei, the average distance to the nearest neighbor $\bar{l}_{nearest}$ is:

$$\bar{l}_{nearest} = \frac{1}{\sqrt{N_{(\infty)}^{2D}}} = \frac{1}{\sqrt{dN(\infty)}} \tag{13}$$

Therefore, the average impingement time $\bar{t}_{imp}$ can be expressed as:

$$\bar{t}_{imp} = \frac{\bar{l}_{nearest}}{2v} = \frac{1}{2v\sqrt{dN(\infty)}} \tag{14}$$

**Numerical calculation and fitting**

The simulated kinetic curves were generated using the Riemann sum method based on the extended KAI model, equation (4). The subinterval width was consistent with the sampling width of experimental acquisition (40 ns) to enable direct comparison. The simulation – experimental curve fitting was conducted with the generalized reduced gradient method to seek the minimal sum of squares error, with variables $\alpha$ and $vN(\infty)^{1/2}$.

**Sample synthesis**

The 250 nm $Al_{0.7}Sc_{0.3}N$ film was deposited on $Pt/TiO_x/SiO_2/Si$ substrate[55] via reactive RF magnetron sputtering using the following growth conditions: 2 mTorr of $Ar/N_2$ (15/5 sccm flow), and a target power density of 6.6 W/cm$^2$ on a 2" diameter $Al_{0.7}Sc_{0.3}$ alloy target (Stanford Advanced Materials). The substrate was rotated and heated to 400 °C during deposition. The base pressure, partial oxygen and water vapor pressure at 400 °C were $< 2 \times 10^{-7}$ torr, $P_{O2} < 2 \times 10^{-8}$ torr and $P_{H2O} < 1 \times 10^{-7}$ torr, respectively.

The 250 nm $Al_{0.94}B_{0.06}N$ film was deposited on W/c-sapphire via magnetron co-sputtering using the following growth conditions: 2 mTorr of $Ar/N_2$ (5/15 sccm flow), with target power densities of 12.5 W/cm$^2$ (pulsed DC) on a 2" diameter Al target (Kurt J. Lesker) and 4.75 W/cm$^2$ (RF) on a 2" diameter BN target (Kurt J. Lesker). The substrate was rotated and heated to 325 °C during deposition. The chamber base pressure at deposition temperature was $< 5 \times 10^{-7}$ torr.

Top Au (100nm)/Ti (5 nm) contacts 50 and 200 μm in diameter were deposited on the $Al_{0.7}Sc_{0.3}N$ and $Al_{0.94}B_{0.06}N$ films via electron beam evaporation through a photolithographically patterned mask.

**Measurements**

Ferroelectric polarization – electric field hysteresis loop measurements were taken with a Precision Multiferroic system from Radiant Technologies. The frequency of the applied triangle excitation voltage was 10 kHz

The polarization switching kinetics were determined with the switching current extracted by positive – up – negative – down (PUND) measurements[56], namely by subtracting the current signal associated with the second same-polarity pulse (including capacitive current and resistive current) from that of the first pulse (including switching current, capacitive current and resistive current). The system was designed to cover < ±200 V and > 200 ns measurement range to observe sample limited ferroelectric switching in sub-microsecond order for 250 nm thick wurtzite ferroelectric films. The data were compared to that for a $LiTaO_3$ single crystal, whose switching kinetics are known to follow the KAI model[14], it was found that the Avrami exponent for the $LiTaO_3$ sample remains 2 across all measurement conditions, consistent with prior studies[40]

Cycling dependence of switching kinetics, piezoelectric properties, and etching effects were compiled after applying different numbers of PUND pulses (Fig. S6, ESI) to otherwise identical electrodes across the same specimen using a non-sequential pattern. Polarization evolution curves were determined from the P-U pulses of each cycle. Piezoelectric coefficients $d_{33}$ were measured using a double beam laser interferometer from aixACCT prior to the P pulse for each PUND cycle. Phosphoric acid etching (85 °C for 10 minutes) and surface SEM observation were carried out in the state prior to P and N pulses for each cycle.

**Author contributions:** KY conceived the study and carried out $Al_{0.7}Sc_{0.3}N$ film synthesis, all the characterization, measurement system development, kinetic model development, simulation, and analysis. JH and JPM synthesized $Al_{0.94}B_{0.06}N$ film. WZ synthesized the $Pt/TiO_x/SiO_2/Si$ substrates. KY, GLB, AZ and STM conducted data interpretation. KY and GLB created effective visualization. GLB and AZ supervised KY. KY, GLB, STM, and AZ composed the original manuscript. All authors contributed to the discussion and reviewing and editing the manuscript.

**Conflicts of interest:** There are no conflicts to declare.

**Acknowledgments:** This work was co-authored by Colorado School of Mines and the National Renewable Energy Laboratory, operated by the Alliance for Sustainable Energy, LLC, for the U.S. Department of Energy (DOE) under Contract No. DE-AC36-08GO28308. Funding was provided by the DARPA Tunable Ferroelectric Nitrides (TUFEN) program (DARPA-PA-19-04-03) as a part of Development and Exploration of FerroElectric Nitride Semiconductors (DEFENSE) project (electrical characterization), by the Office of Science (SC), Office of Basic Energy Sciences (BES) as part of the Early Career Award "Kinetic Synthesis of Metastable Nitrides" (AlScN material synthesis), and by the National Science Foundation under Grant No. DMR-2119281 (data analysis, and manuscript preparation). Development of the $Al_{1-x}B_xN$ film growth methods was supported by grants DARPA grants HR0011-20-9-0047 and W911NF-20-20274, while the film deposition was supported as part of the center for 3D Ferroelectric Microelectronics (3DFeM), an Energy Frontier Research Center funded by the U.S. Department of Energy (DOE), Office of Science, Basic Energy Sciences under Award Number DE-SC0021118. We also thank Dr. Kevin Talley of Mines and NREL and Danny Drury of Mines,

# Supporting Information for

**Anomalously Abrupt Switching of Ferroelectric Wurtzites**

Keisuke Yazawa[1,2]*, John Hayden[3], Jon-Paul Maria[3], Wanlin Zhu[3], Susan Trolier-McKinstry[3], Andriy Zakutayev[1], Geoff L. Brennecka[2]*

*Corresponding authors: Email: Keisuke.Yazawa@nrel.gov; geoff.brennecka@mines.edu

**Supplementary Figures**

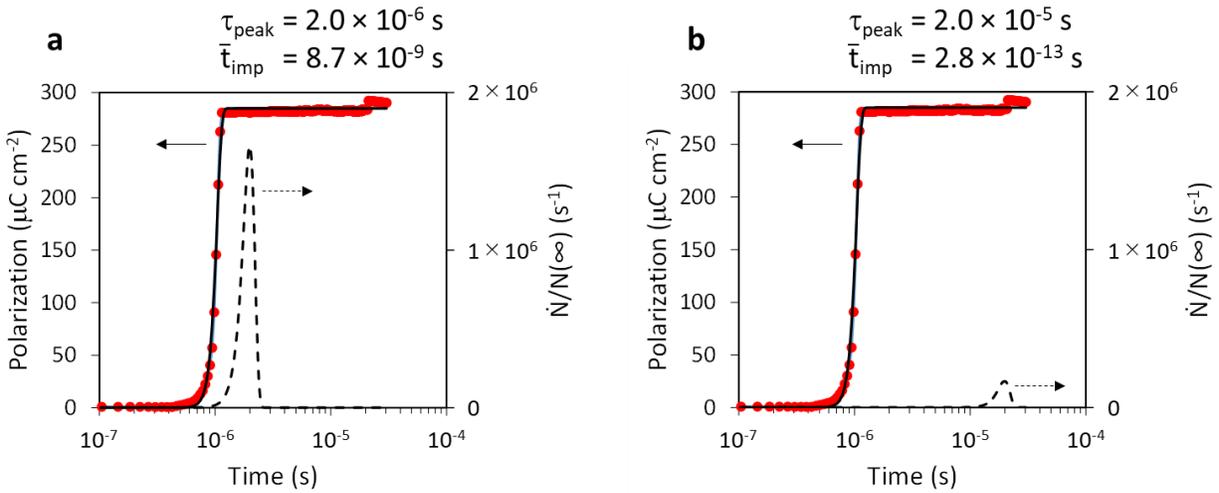

**Fig. S1. Fitting results with the extended KAI model. Black line is the simulated curve, and red circles are the experimental results for Al$_{0.94}$B$_{0.06}$N (a)** selected faster nucleation rate ($\tau_{peak}$ = 2 × 10$^{-6}$ s) and **(b)** selected slower nucleation rate ($\tau_{peak}$ = 2 × 10$^{-5}$ s). Both the simulated curve fit well with the experimental results with the optimal domain wall velocities or average impingement time; impingement time of the fast nucleation rate simulation is faster than that of slow nucleation rate simulation. Thus, the $\tau_{peak}$ and $\bar{t}_{imp}$ values using the model have a fittable range.

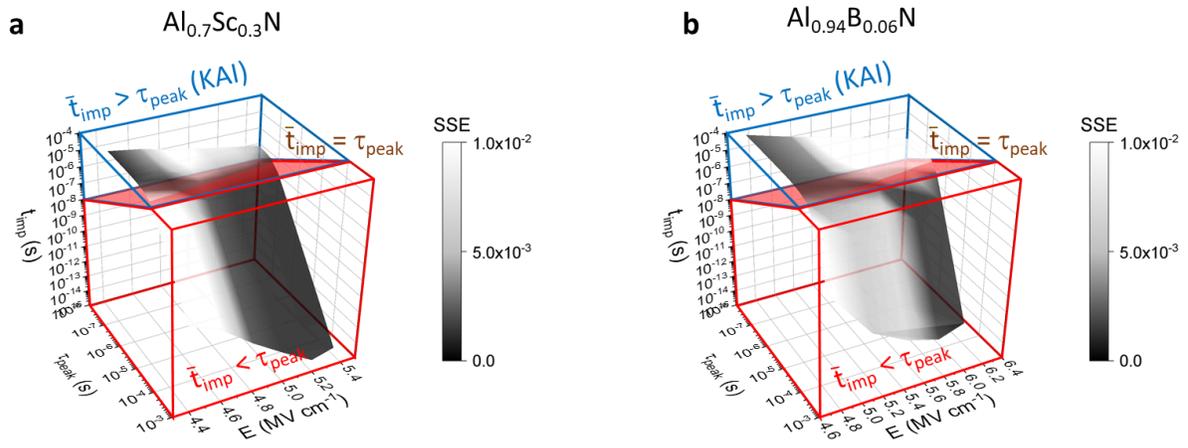

**Fig. S2. Fitting quality map as a function of fitting parameters. The sum of squares of error (SSE) for fitting the $t_{peak} - \bar{t}_{imp} - E$ surface is represented with the depth of gray scale, with darker regions representing better fits.** (**a**) Fitted $\tau_{peak} - \bar{t}_{imp} - E$ surface shaded according to SSE for $Al_{0.7}Sc_{0.3}N$ and (**b**) for $Al_{0.94}B_{0.06}N$. The red-colored plane (seen nearly edge-on in both plots) represents $\tau_{peak} = \bar{t}_{imp}$, and low SSE (well fitted) conditions pass across the plane with varying electric field for both materials.

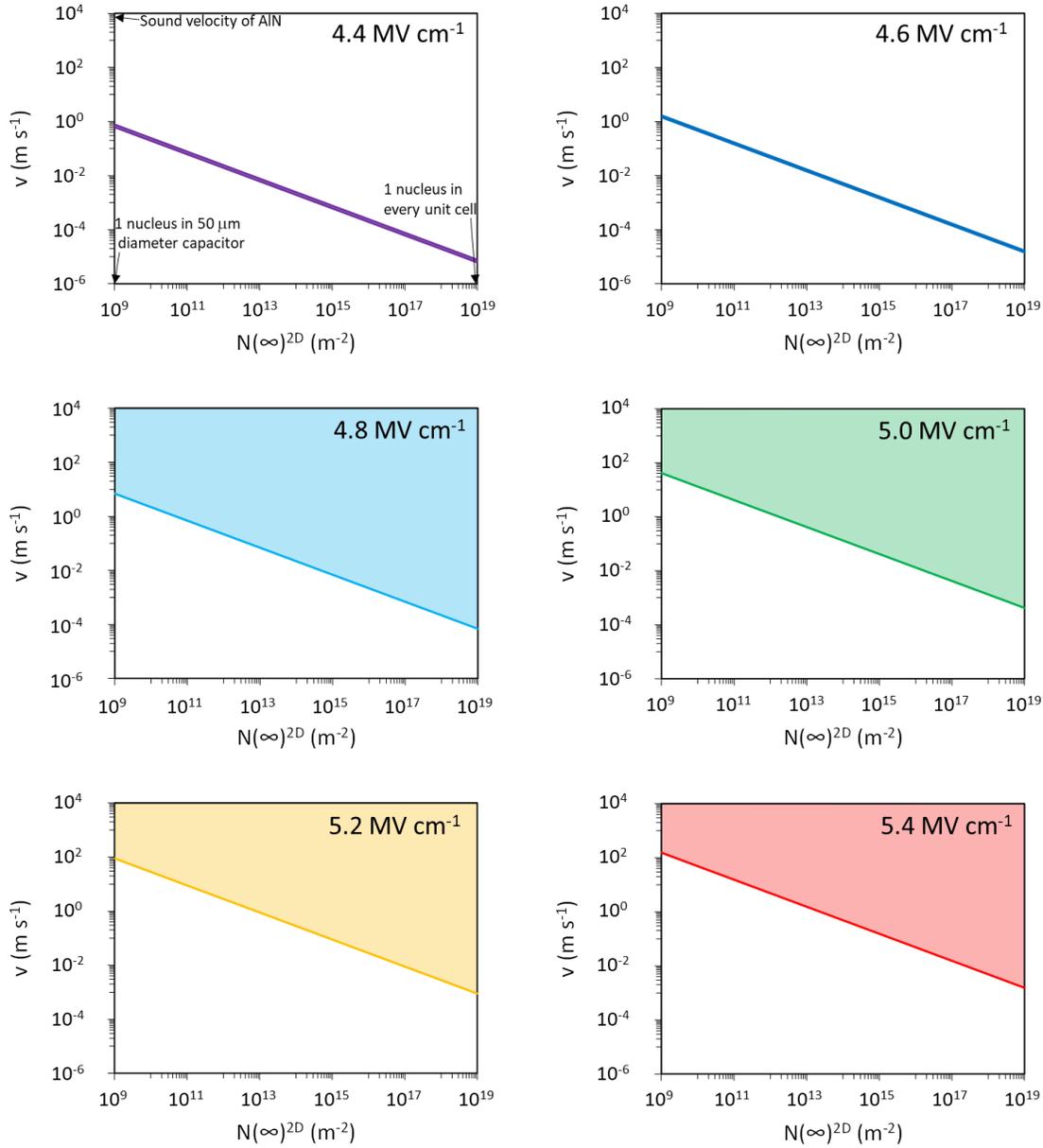

**Fig. S3.** Domain wall velocity of $Al_{0.7}Sc_{0.3}N$ as a function of saturated nucleation densities $N(\infty)^{2D}$ for various electric fields, 4.4 MV cm$^{-1}$ ~ 5.4 MV cm$^{-1}$. The shaded region represents the possible range due to the range of the fitted $\bar{t}_{imp}$ or $vN(\infty)^{1/2}$. The saturated nucleation densities limited to the practical range from one nucleus for the measured device size to one nucleus for each unit lattice. The velocity range shown is up to the sound velocity of AlN. Physically reasonable velocities for the domain wall exist independent of the nucleation density assumed.

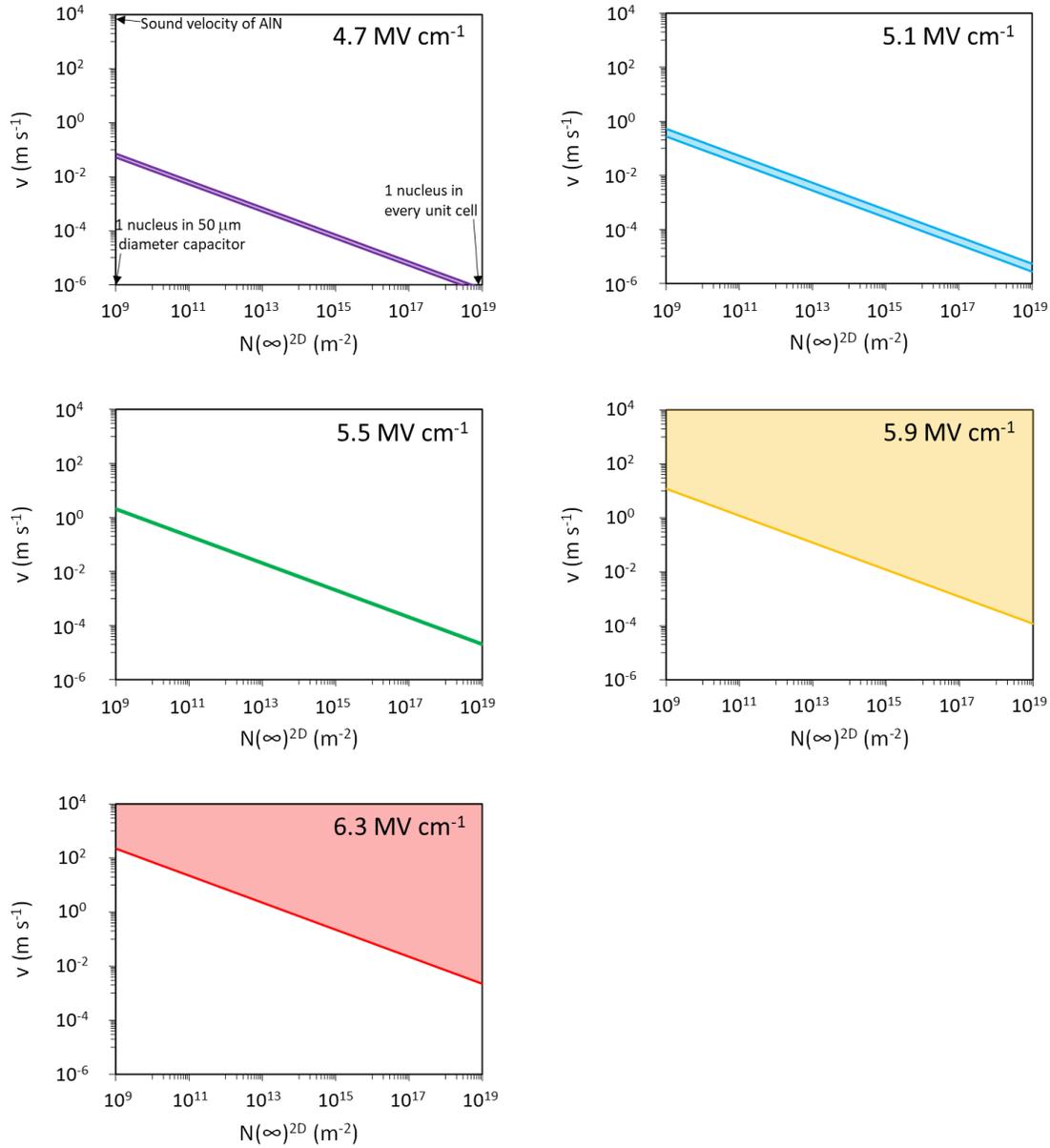

**Fig. S4.** Domain wall velocity of $Al_{0.94}Sc_{0.06}N$ as a function of saturated nucleation densities $N(\infty)^{2D}$ for various electric fields, 4.7 MV cm$^{-1}$ ~ 6.3 MV cm$^{-1}$. The shaded region represents the possible range arising from the fitted $\bar{t}_{imp}$ or $vN(\infty)^{1/2}$. The saturated nucleation densities shown are limited to the range from one nucleus per measured capacitor to one nucleus for each unit lattice. The velocity is plotted up to the sound velocity of AlN. Physically reasonable domain wall velocities exist for all plausible nucleation densities.

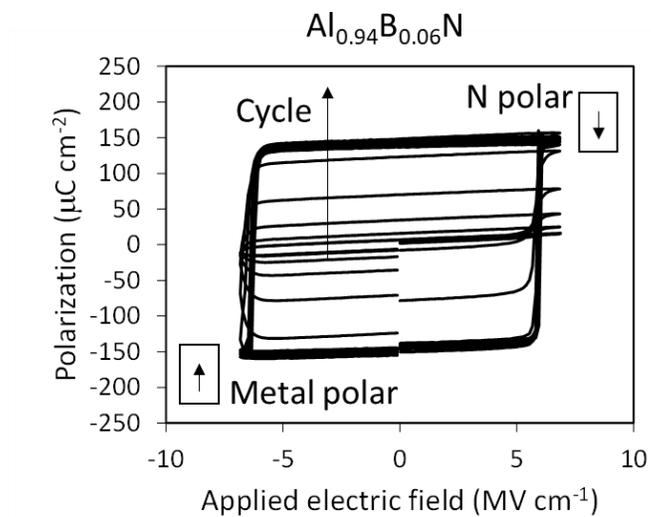

**Fig. S5.** P-E hysteresis loop with wakeup behavior on wurtzite Al$_{0.94}$B$_{0.06}$N film, showing the significant wakeup behavior. The hysteresis loops are taken under a repeated triangular excitation field (6.9 MV cm$^{-1}$ and 10 kHz).

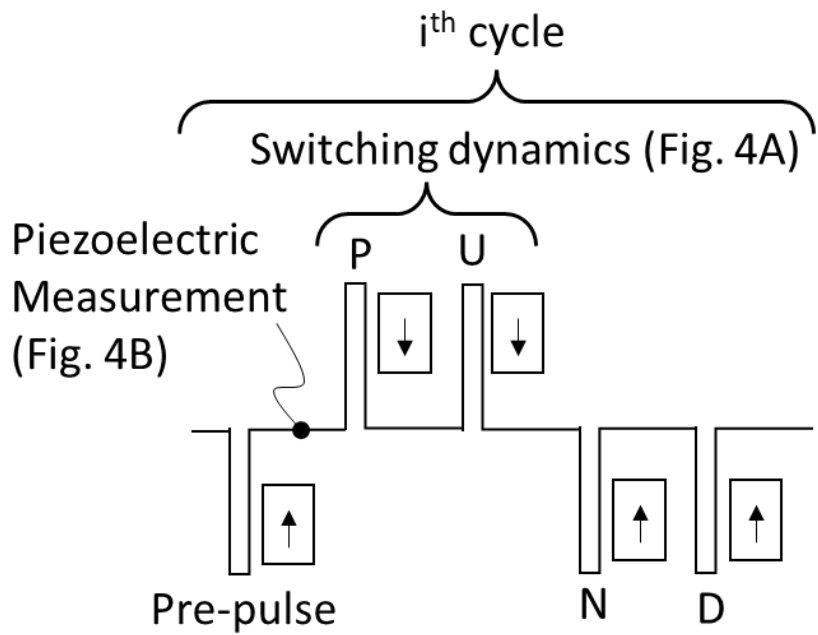

**Fig. S6.** Pulse sequence for cycling dependence of switching dynamics and piezoelectric property. The piezoelectric property taken prior to P pulse represents state just prior to the switching dynamics measurements.

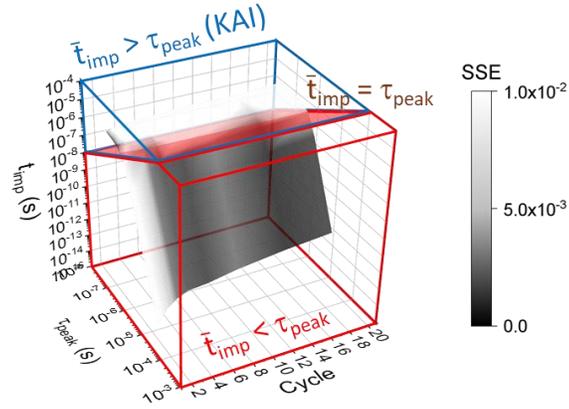

**Fig. S7.** Fitting quality map as a function of fitting parameters. The sum of squares of error (SSE) for fitting the $t_{peak} - \bar{t}_{imp} -$ cycle surface is represented with the depth of gray scale for $Al_{0.94}B_{0.06}N$, with darker regions representing better fits. The red-colored plane (seen nearly edge-on in both plots) represents $\tau_{peak} = \bar{t}_{imp}$, and low SSE (well fitted) conditions pass across the plane with varying number of cycles.

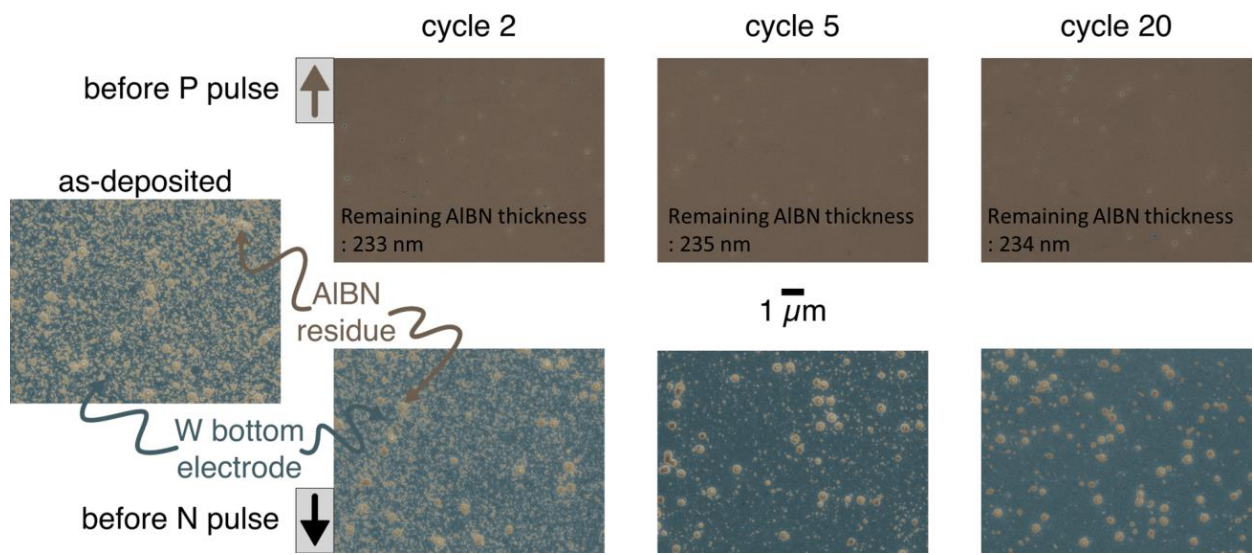

**Fig. S8.** SEM images of $Al_{0.94}B_{0.06}N$ films after phosphoric acid etching (see Methods). Before P pulses, slowly - etched metal polar $Al_{0.94}B_{0.06}N$ film remains ~235 nm thick (~15 nm etched) regardless of number of cycles > 2, which indicates that there is no significant difference in the polarity at the top electrode/film interface as a function of the number of cycles. As deposited and before N pulses, fast – etched nitrogen polar $Al_{0.94}B_{0.06}N$ film is etched away and the underlying W electrode is exposed. The $Al_{0.94}B_{0.06}N$ residue decreases with increasing cycle number.